**Three-dimensional microtomographic imaging of human brain cortex**

Ryuta Mizutani[a,*], Akihisa Takeuchi[b], Kentaro Uesugi[b], Masami Ohyama[a], Susumu Takekoshi[c], R. Yoshiyuki Osamura[c], Yoshio Suzuki[b]

[a]Department of Applied Biochemistry, School of Engineering, Tokai University, Kitakaname 1117, Hiratsuka, Kanagawa 259-1292, Japan

[b]Research and Utilization Division, JASRI/SPring-8, Kouto 1-1-1, Sayo, Hyogo 679-5198, Japan

[c]Department of Pathology, Tokai University School of Medicine, Shimokasuya 143, Isehara, Kanagawa 259-1193, Japan

*Corresponding author, Fax: +81-463-50-2506. E-mail address: ryuta@keyaki.cc.u-tokai.ac.jp (R. Mizutani).

**Abstract**

This paper describes an x-ray microtomographic technique for imaging the three-dimensional structure of the human cerebral cortex. Neurons in the brain constitute a neural circuit as a three-dimensional network. The brain tissue is composed of light elements that give little contrast in a hard x-ray transmission image. The contrast was enhanced by staining neural cells with metal compounds. The obtained structure revealed the microarchitecture of the gray and white matter regions of the frontal cortex, which is responsible for the higher brain functions.

**1. Introduction**

The neuronal network of the central nervous system (CNS) is composed of a huge number of neurons, which are responsible for the cerebral function. The functional mechanism of the brain can therefore be revealed by visualizing the neurons constituting the three-dimensional structure. Confocal optical microscopy is the primary method for visualizing three-dimensional structures of biological systems (Conchello and Lichtman 2005). However, absorbance at emission or excitation wavelengths interferes with the detection of fluorescence elicited from the internal architecture. Thus, confocal microscopy is mainly used for imaging sectioned samples labelled with highly selective probes. Reconstruction of the entire three-dimensional structure from such samples entails defining the precise spatial relationships of each image. Individual variations of the structure of each preparation create constitutive difficulties in superposing multiple images. Optical computed tomography (CT) and ultramicroscopy have recently been applied to the three-dimensional analysis of biological tissues (Alanentalo et al., 2007; Dodt et al., 2007), although the internal structure of opaque samples cannot be observed by using visible light. While magnetic resonance imaging provides a non-invasive mapping of neural activity (Yu et al., 2005), cellular structures are not visualized due to the limited spatial resolution.

In contrast, the transparency of biological tissue to hard x-rays enables radiographic analysis of the microstructure. However, biological tissues are mainly composed of light elements, which produce little contrast in a hard x-ray transmission image. In clinical diagnosis, luminal structures of a living body are visualized by using x-ray contrast media. These contrast media contain heavy-atom elements that absorb x-rays efficiently. We have recently shown that the neuronal structure of the *Drosophila* larvae CNS can be visualized by contrasting every neuron with the metal impregnation method (Mizutani et al., 2007). Recent application of synchrotron radiation to high-resolution CT (micro CT) has resolved three-dimensional structures at the micrometer (Uesugi et al., 2001) to submicrometer scales (Takeuchi et al., 2002). In conjunction with the metal microcontrasting, micro-CT analysis has revealed the entire three-dimensional structure of the *Drosophila* larvae CNS (Mizutani et al., 2007).

The microcontrasting CT analysis can be applied to any biological tissues. A number of metal impregnation methods have been developed for the optical observation of biological tissues. Golgi silver impregnation (Valverde ,1970) is a conventional staining method for the optical observation of neural cells. Here, we report a micro-CT analysis of the human cerebral cortex stained by the Golgi impregnation.

**2. Results**

Reconstructed images revealed the three-dimensional structure of the cerebral cortex. The structure of a columnar sample dissected from the frontal cortex of the human brain is shown in Fig. 1A–E. Since the viewing field of this micro-CT analysis is limited to a cylindrical region 1.00 mm in diameter x 0.65 mm in height, data sets of the columnar sample with an approximate diameter of 0.7 mm and height of 2.3 mm were taken in four batches. Each data set was recorded by displacing the sample 550 µm along the columnar axis. It took about 30 min to acquire each dataset, though the imaging time can be shortened by reducing the viewing field or the number of radiographs if the faster acquisition is required. The obtained radiographs for four data sets were subjected to reconstruction calculation. An example of the structure is shown in Fig. 1B, which indicates the neuronal structure surrounded by the glass wall of the capillary tube. To determine the precise positional relationship of the four reconstructed structures, we superposed each end of the three-dimensional structures by minimizing the root-mean-square difference of voxel density using the program RMSD (Nakano et al., 2006). The whole image was built up by stacking these data sets. Then, voxels corresponding to the glass wall were removed for easier recognition of the three-dimensional structure of the nerve tissue. The tissue components stained with the metal dye were obviously distinguished from the unstained surroundings (Fig. 1C).

The cerebral hemispheres consist of gray matter and white matter. The upper two-thirds of the columnar sample used in the present study correspond to the gray matter. Histologically, neurons of the gray matter of the frontal cortex are arranged into six layers, called the molecular layer (layer I), external granular layer (II), external pyramidal layer (III), internal granular layer (IV), internal pyramidal layer (V), and multiform layer (VI). The external and internal pyramidal layers are characterized by pyramidal neurons. In the middle of the three-dimensional structure shown in Fig. 1C, a group of cortical pyramidal cells was observed. This pyramidal cell layer can be assigned as the internal pyramidal layer. Apical dendrites from the apex of the pyramidal cells were observed as long fibriform structures that projected into the upper layer of the cerebral cortex (Fig. 1D and Supplementary Movie 1A). Smaller pyramidal cells were found in the upper layer, which corresponds to the external pyramidal layer. In the region of the multiform layer, pyramidal and granular cells were seen. A magnified image of pyramidal cells in the internal pyramidal layer is shown in Fig. 1E and Supplementary Movie 1B. Axons and dendrites arising from somas were visualized as a network structure. In the cutaway section at the top of Fig. 1E, a low-density region can be seen in the interior of a soma, indicating that the micro-CT analysis can reveal the intracellular microstructure. This microstructure can be considered to be the cellular nucleus.

Luminal structures can be seen beneath the internal pyramidal layer in Fig. 1C. These structures were assigned as blood vessels stained by the Golgi impregnation because the cut end of the upstream vessel was exposed at the sample surface. Such luminal structures were mainly observed in the white matter. Figure 2A shows the structure of the larger prismatic sample analyzed at 8-µm resolution, indicating that luminal networks are seen in the middle part corresponding to the white matter. The structure of the prismatic sample with dimensions of 1.4 mm x 1.4 mm x 8.3 mm was built up from three data sets, for which the viewing field corresponded to a cylindrical region 5.5 mm in diameter x 3.6 mm in height. The gray matter is in the upper layer and in the lower-left corner. These regions indicate an anatomical hallmark characterized by localization of dense granules. These granules can be assigned as somas. In Fig. 2B, the layer formed by pyramidal cells can obviously be recognized.

## 3. Discussion

Microstructures of biological tissues have generally been characterized from thin histological slices. The mechanical stress introduced by microtomy procedures results in physical distortion of the tissue structure, making it difficult to reconstruct the three-dimensional structure from slice images. In contrast, the transparency of biological tissue to hard x-rays enables the micro-CT analysis of the entire structure. The main advantages of such microcontrasting CT analysis compared with conventional histology are the three-dimensional nature of the obtained structure and the exclusion of artificial deformation originating from sectioning preparation. The three-dimensional microarchitecture of the human brain visualized in this study was composed of a large number of neurons, which are responsible for the cerebral function.

The three-dimensional microstructures obtained by micro-CT analysis could be utilized in histological diagnosis. In the present histology, two-dimensional optical microscopic images are the primary information for diagnosis. Such histological images provide hardly any information about the three-dimensional nature of clinical samples. Radiographic diagnosis of visceral lesions has been innovated by medical application of CT imaging. Similarly, micro-CT analysis along with metal microcontrasting of clinical samples can provide novel pathological information that cannot be visualized in two-dimensional images. Although the present analysis was performed on post-mortem tissue of the human brain, biopsy samples of any organ could be subjected to the microcontrasting CT analysis. The three-dimensional structure of such a sample provides the entire image of the biopsied tissue, from which any arbitrary two-dimensional sections could be generated.

The three-dimensional transcriptome map of the entire mouse brain has been reported (Lein et al., 2007). It was pointed out that an important future goal is to determine the relationship between transcriptional profiles and cellular phenotypes by correlating gene expression with more salient characteristics such as morphology. Although microstructure analysis of the entire brain is necessary for the elucidation of the brain function, the viewing field provided by micro-CT analysis cannot cover the whole brain. The size limitation of the present experiment setup should be overcome by modifying the detection strategy to cover a larger sample area.

The elaborate neuronal network visualized in the three-dimensional structure is essential for cerebral functions. Therefore, we want to analyze the entire network to reveal the functional mechanism of the brain. The brain model should be built up as a circuit so as to reveal its functions, so it is essential to trace neuronal tracts in order to understand the functional mechanism of the brain. It is difficult in principle to trace the neuronal tracts manually because the three-dimensional tomographic structure of the brain is much more complicated than selectively labelled fluorescence images, which are confined to specific neurons. The higher-resolution structure determined by zone plate optics (Takeuchi et al., 2002; Mizutani et al., 2007) would enable automatic neuron tracing on a three-dimensional image.

The Golgi impregnation stains only some neural cells. For the visualization of the entire neuronal structure, metal microcontrasting techniques should be further developed so that they can stain every neuron in the whole brain. Selective staining using neuronal marker antibodies has been reported for histological analysis (Evers and Uylings, 1997). Gold and other metal/antibody conjugates used as electron-dense markers, especially in electron microscopic imaging (Zuber et al., 2005), can be applied to the present CT analysis.

The wavelength of x-rays is several hundred times-shorter than the visible light used in optical microscopy. Therefore, the diffraction-limited resolution of x-ray microscopy is much higher than that of optical microscopy. We have reported a radiographic analysis of nerve tissue at 160-nm resolution (Mizutani et al., 2007) visualizing the neuronal structures. Micro-CT analysis at this higher resolution enables functional analysis of the neuronal network from the three-dimensional microstructure of the nerve tissue.

## 4. Experimental procedures

4.1. Tissue samples

Frontal cortex of normal brain tissue (44 years old, male) was dissected at autopsy and fixed with 10% formaldehyde for 7 days. Anatomical analysis found no abnormality in the brain tissue. The tissue sample was further dissected into 5-mm tissue cubes and subjected to Golgi staining (Valverde, 1970) by the following procedure. The tissues were washed for 5 min in a solution containing 2.5% potassium dichromate and 4% formaldehyde and then incubated at 25ºC for 7 days in a solution containing 2.5% potassium dichromate. After being blotted with filter paper, the tissues were further incubated at 25ºC for 48 hrs in 0.75% silver nitrate. Residual silver nitrate was washed away with water. The tissues were then sequentially immersed in ethanol, *n*-butylglycidyl ether, and Petropoxy 154 epoxy resin (Burnham Petrographics, ID). Human samples were obtained with written consent, using protocols approved by the Clinical Research Review Board of Tokai University Hospital.

Columnar samples were dissected from the stained tissues using a borosilicate glass capillary with an outer diameter of 0.7 mm (W. Müller Glas, Germany). The capillaries carrying the samples were incubated at 90ºC for 16 hours to rigidify the epoxy resin, and then used for the micro-CT analysis.

The larger prismatic samples with dimensions of 1.4 mm x 1.4 mm x 8.3 mm were prepared using razor blades from the longer tissue blocks (5 x 5 x 10 mm$^3$ approximately). The samples were placed in silicone tubing filled with epoxy resin. The inner diameter of the tubing was 3 mm. The samples were incubated at 90ºC for 16 hours, and then the silicone tubing was removed. The obtained resin pellets were used for the micro-CT analysis.

4.2. Micro CT

The columnar samples were analyzed at the BL20XU beamline (Suzuki et al., 2004) of SPring-8, Japan. The samples were mounted on the goniometer head of the microtomograph, using a brass fitting designed for the glass capillary sample. Transmission radiographs were recorded with a CCD-based x-ray imaging detector (AA50 and C4880-41S, Hamamatsu Photonics, Japan) using 12.0-keV monochromatic x-rays. The field of view and effective pixel size of the image detector were 1.00 mm x 0.65 mm and 0.50 μm x 0.50 μm, respectively. A total of 1800 images were acquired with a rotation step of 0.10º and an exposure time of 300 ms. The spatial resolution of the three-dimensional structure was estimated to be 1.0 μm in each direction.

The larger prismatic samples were analyzed at the BL20B2 beamline (Goto et al., 2001) of SPring-8. The sample pellets were mounted on the goniometer head of the microtomograph, using adhesive tapes. Transmission radiographs were recorded using a CCD-based x-ray imaging detector (AA40P and C4880-41S, Hamamatsu Photonics) and 12.0-keV x-rays. The field of view and effective pixel size of the detector were 5.5 mm x 3.6 mm and 2.75 μm x 2.75 μm, respectively. Each image was acquired with a rotation step of 0.10º and an exposure time of 600 ms. The spatial resolution of the three-dimensional structure was estimated to be 8 μm.

The convolution back projection method using a Chesler-type filter was used for tomographic reconstruction (Uesugi et al., 2001). The reconstruction calculation takes about 65 hours on a Windows PC equipped with the Core 2 processor operating at 2.1 GHz. Reconstructed images were further processed using the program suite "slice" (Nakano et al., 2006). Volume-rendered figures of the obtained three-dimensional structures were produced using the program VG Studio MAX (Volume Graphics, Germany). CT densities were rendered using the scatter HQ algorithm.

The data acquisition conditions are summarized in Table 1.

**Acknowledgements**

We thank Dr. Junko Matsuda (Institute of Glycotechnology, Tokai University) for discussion and Noboru Kawabe (Teaching and Research Support Center, Tokai University School of Medicine) for sectioning tissue. The synchrotron radiation experiments were performed at SPring-8 with the approval of the Japan Synchrotron Radiation Research Institute

# References

Alanentalo, T., Asayesh, A., Morrison, H., Lorén, C.E., Holmberg, D., Sharpe, J., Ahlgren, U., 2007. Tomographic molecular imaging and 3D quantification within adult mouse organs. Nat. Methods 4, 31-33.

Conchello, J.-A., Lichtman, J.W., 2005. Optical sectioning microscopy. Nat. Methods 2, 920-931.

Dodt, H.-U., Leischner, U., Schierloh, A., Jährling, N., Mauch, C.P., Deininger, K., Deussing, J.M., Eder, M., Zieglgänsberger, W., Becker, K., 2007. Ultramicroscopy: three-dimensional visualization of neuronal networks in the whole mouse brain. Nat. Methods 4, 331-336.

Evers, P., Uylings, H.B.M., 1997. An optimal antigen retrieval method suitable for different antibodies on human brain tissue stored for several years in formaldehyde fixative. J. Neurosci. Methods 72, 197-207.

Goto, S., Takeshita, K., Suzuki, Y., Ohashi, H., Asano, Y., Kimura, H., Matsushita, T., Yagi, N., Isshiki, M., Yamazaki, H., Yoneda, Y., Umetani, K., Ishikawa, T., 2001. Construction and commissioning of a 215-m-long beamline at SPring-8. Nucl. Instrum. and Meth. A467-468, 682-685.

Lein, E.S. *et al.*, 2007. Genome-wide atlas of gene expression in the adult mouse brain. Nature 445, 168-176.

Mizutani, R., Takeuchi, A., Hara, T., Uesugi, K., Suzuki, Y., 2007. Computed tomography imaging of the neuronal structure of *Drosophila* brain. J. Synchrotron Radiat.14, 282-287.

Nakano, T., Tsuchiyama, A., Uesugi, K., Uesugi, M., Shinohara, K., 2006. "Slice" -softwares for basic 3-D analysis-, http://www-bl20.spring8.or.jp/slice/, Japan Synchrotron Radiation Research Institute.

Suzuki, Y., Uesugi, K., Takimoto, N., Fukui, T., Aoyama, K., Takeuchi, A., Takano, H., Yagi, N., Mochizuki, T., Goto, S., Takeshita, K., Takahashi, S., Ohashi, H., Furukawa, Y., Ohata, T., Matsushita, T., Ishizawa, Y., Yamazaki, H., Yabashi, M., Tanaka, T., Kitamura, H., Ishikawa, T., 2004. Construction and commissioning of a 248 m-long beamline with x-ray undulator light source. AIP Conf. Proc. 705, 344-347.

Takeuchi, A., Uesugi, K., Takano, H., Suzuki, Y., 2002. Submicrometer-resolution three-dimensional imaging with hard x-ray imaging microtomography. Rev. Sci. Instrum. 73, 4246-4249.

Uesugi, K., Suzuki, Y., Yagi, N., Tsuchiyama, A., Nakano, T., 2001. Development of high spatial resolution X-ray CT system at BL47XU in SPring-8. Nucl. Instrum. and Meth. A467-468, 853-856.

Yu, X., Wadghiri, Y.Z., Sanes, D.H., Turnbull, D.H., 2005. In vivo auditory brain mapping in mice with Mn-enhanced MRI. Nat. Neurosci. 8, 961-968.

Valverde, F., 1970. The Golgi method. In Contemporary Research Methods in Neuroanatomy, W.J.H. Nauta and S.O.E. Ebbesson, ed. Springer-Verlag, New York, pp. 12-31.

Zuber, C., Fan, J., Guhl, B., Roth, J., 2005. Applications of immunogold labeling in ultrastructural pathology. Ultrastruct. Pathol. 29, 319-330.

**Figure legends**

Fig. 1 - Structures of the columnar sample of the human frontal cortex. In panels B-E, three-dimensional structures of the columnar sample are shown as stereo representations. The brain surface is to the top. (A) Optical microscope image of a 15-μm section of the sample subjected to micro-CT analysis. (B) Cutout from the reconstructed three-dimensional structure. The neuronal structure is surrounded by the glass capillary wall. The cutout length along the capillary axis is 200 μm. CT densities are rendered from low linear-absorption-coefficient level (gray, 12.4 cm$^{-1}$) to high level (white, 211 cm$^{-1}$). (C) Overall image of the columnar sample showing the layer structure of nerve cells. Regions of external granular layer (II), external pyramidal layer (III), internal granular layer (IV), internal pyramidal layer (V), and multiform layer (VI) are indicated. An arrow head indicates a cut end of the blood vessel network beneath the multiform layer. CT densities are rendered at 16.5 cm$^{-1}$. (D) Structure around the internal pyramidal layer. An internal pyramidal cell (IP) and its apical dendrite (AD) is labeled and magnified in panel E. CT densities are rendered at 14 cm$^{-1}$. (E) Magnified image of internal pyramidal cells in a 200 x 200 x 355 μm$^3$ rectangular prism. CT densities are rendered from low level (gray, 10 cm$^{-1}$) to high level (white, 50 cm$^{-1}$). The low-density region in the interior of the soma is indicated by an arrow head. Scale bars, 40 μm (A, E) and 100 μm (B-D).

Fig. 2 - Microtomographic images of the prismatic sample of the human frontal cortex. The brain surface is to the top. (A) Overall structure. Gray matter characterized by the soma distribution is indicated by arrows. Blood vessels are localized in the white matter region. CT densities are rendered at a linear absorption coefficient of 24.8 cm$^{-1}$. (B) Structure of the gray matter. Pyramidal cell layer is indicated by arrows. CT densities are rendered at 37.2 cm$^{-1}$. Scale bars, 250 μm.

**Supplementary Movie captions**

Movie 1 - Microtomographic images of the columnar sample of the human frontal cortex. Three-dimensional structures of the columnar sample are shown as surface models. Movies were produced using the program suite "slice". The brain surface is to the top. (A) Structure around the internal pyramidal layer. CT densities are contoured at 33 cm$^{-1}$. (B) Magnified image of internal pyramidal cells in a 200 x 200 x 355 μm$^3$ rectangular prism. CT densities are contoured at 30 cm$^{-1}$.

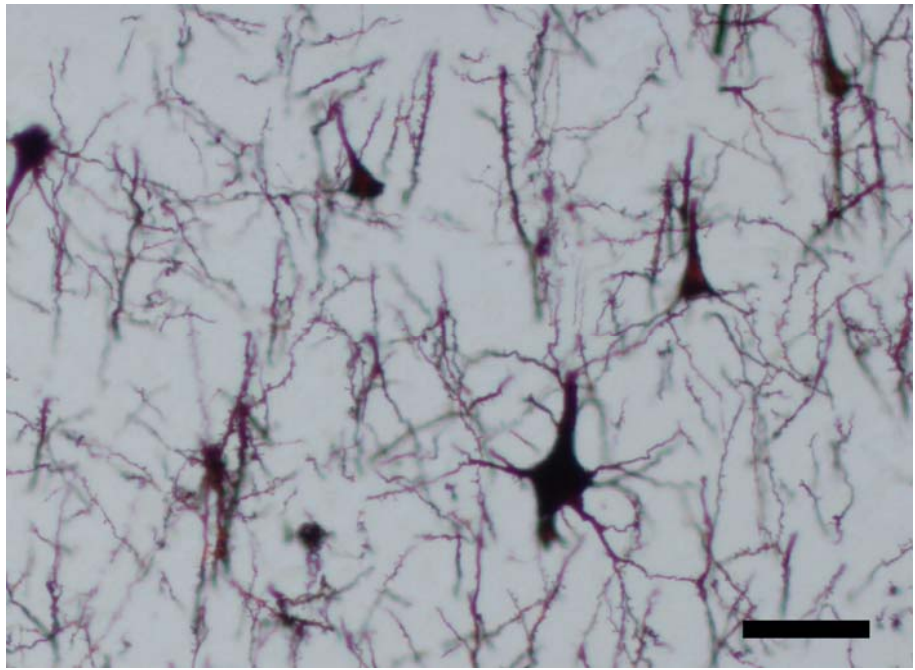

Fig. 1A

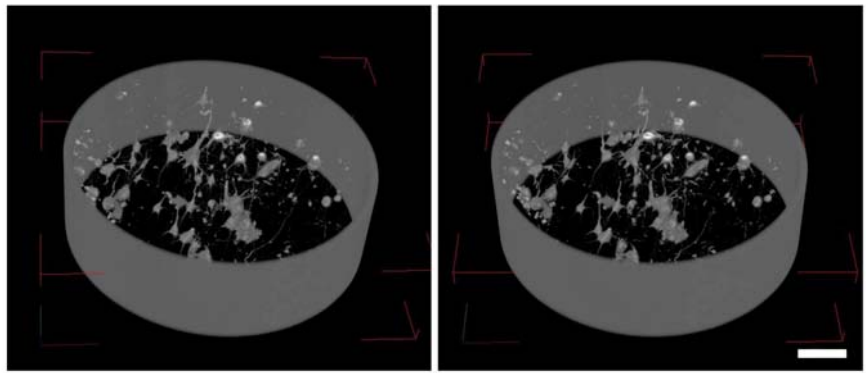

Fig. 1B (actual size)

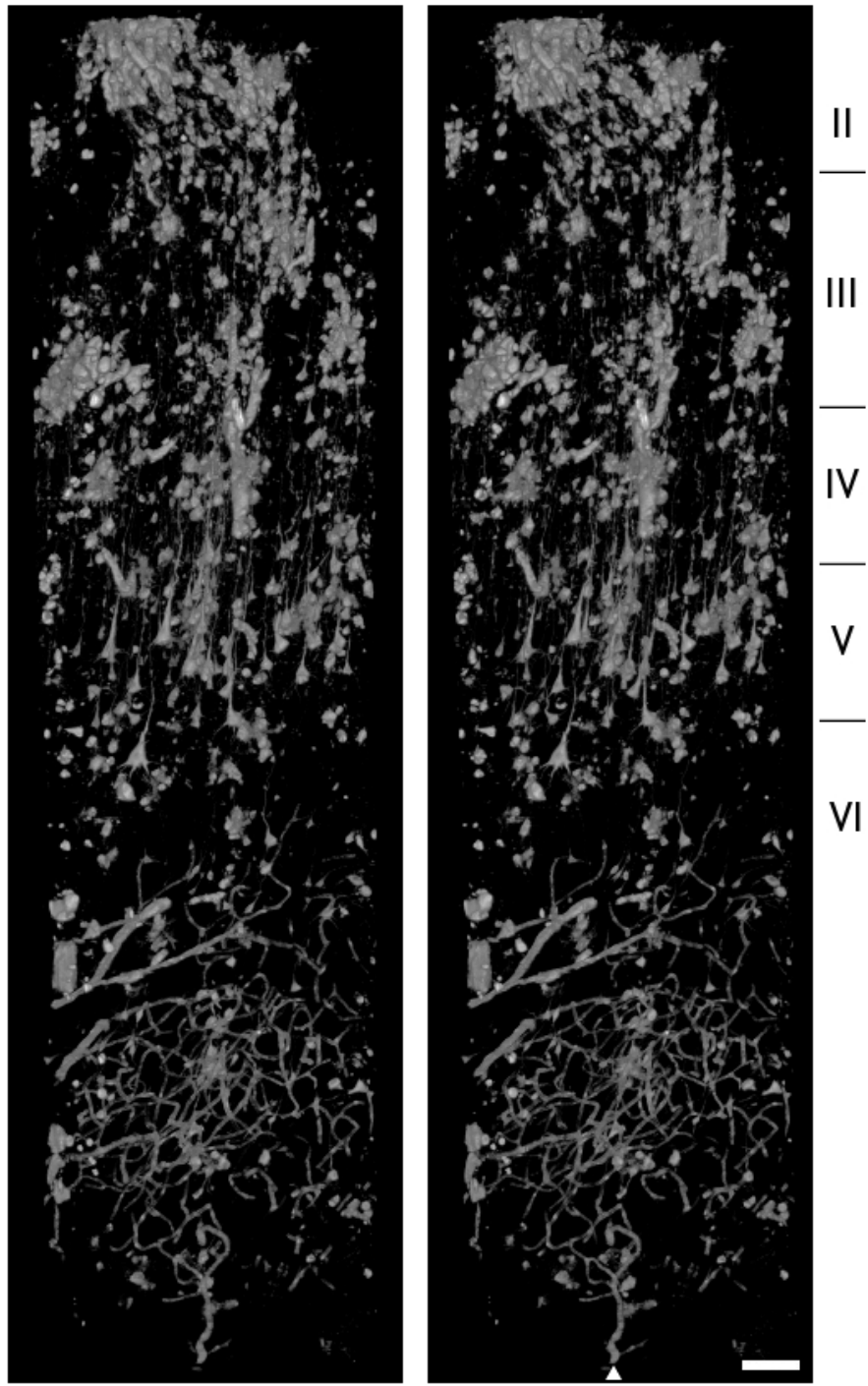

Fig. 1C (actual size)

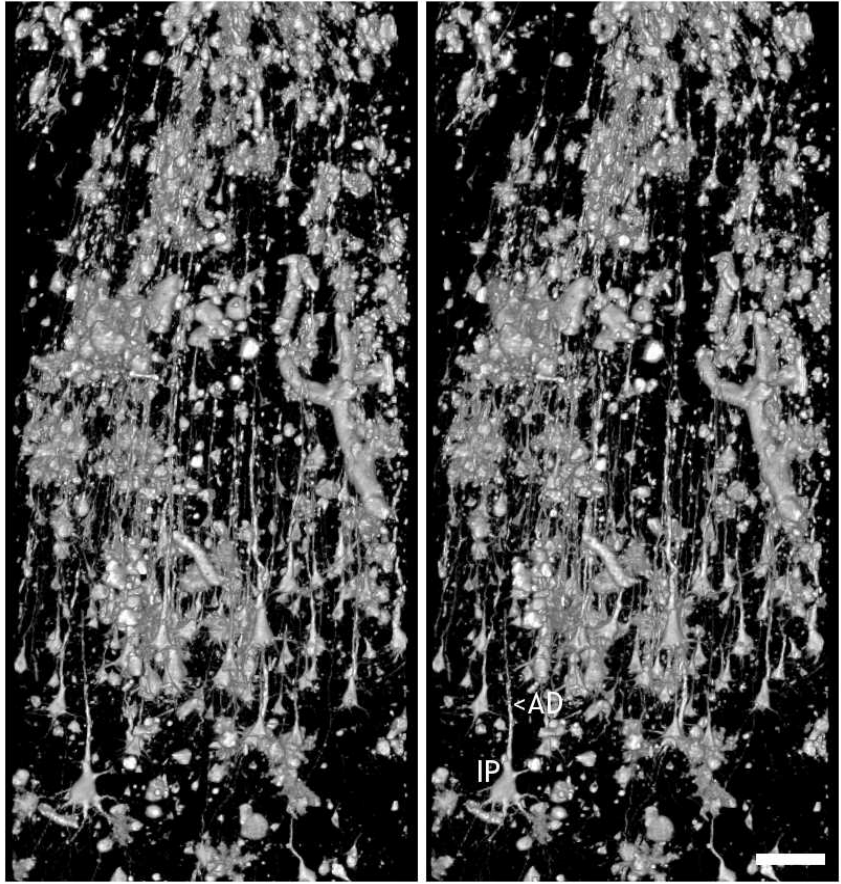

Fig. 1D (actual size)

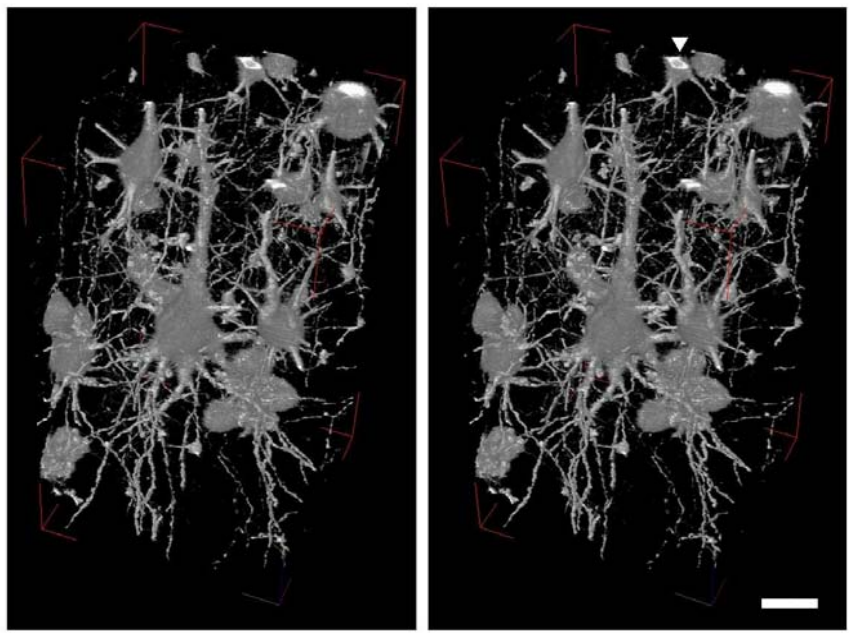

Fig. 1E (actual size)

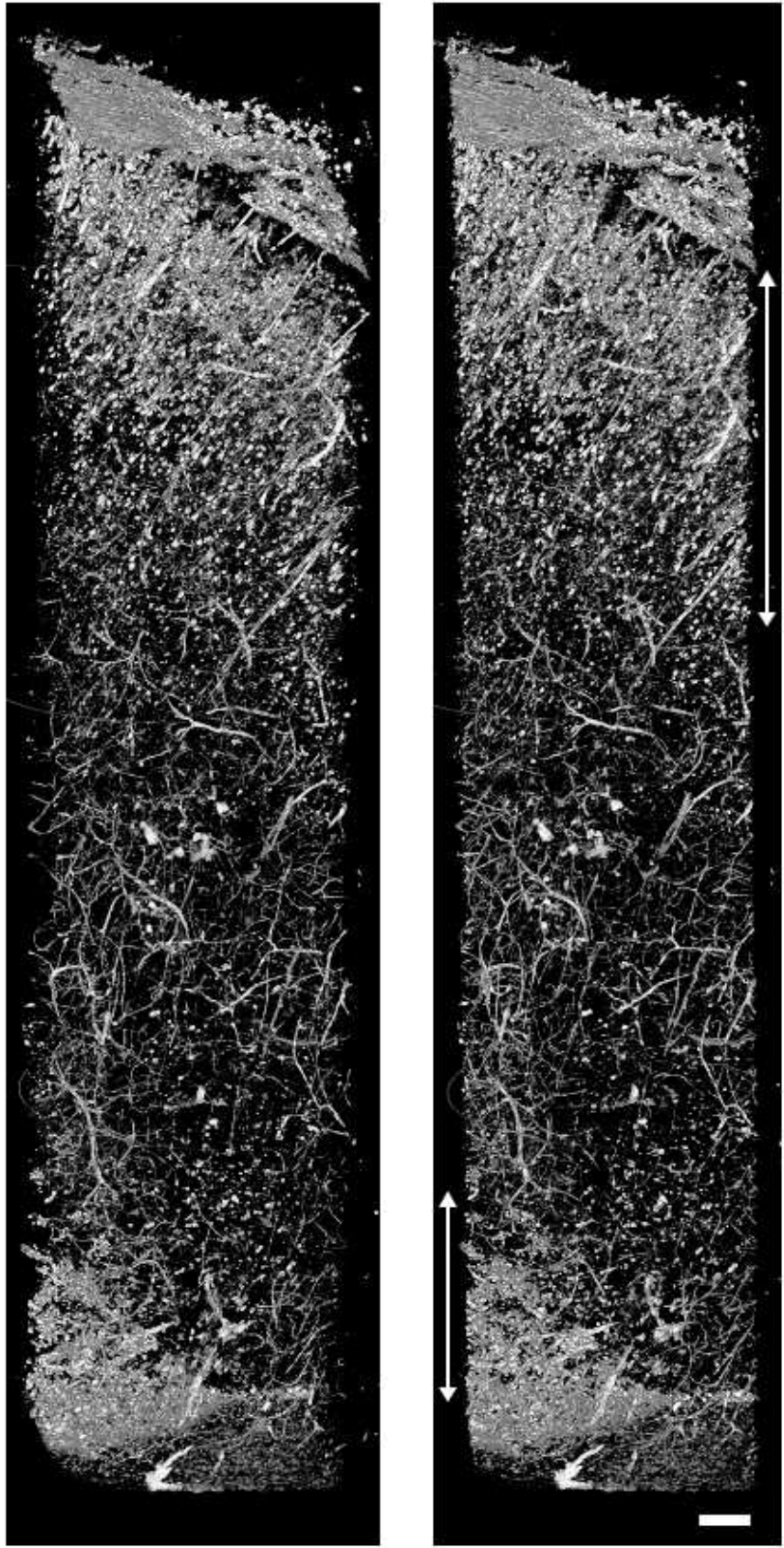

Fig. 2A (actual size)

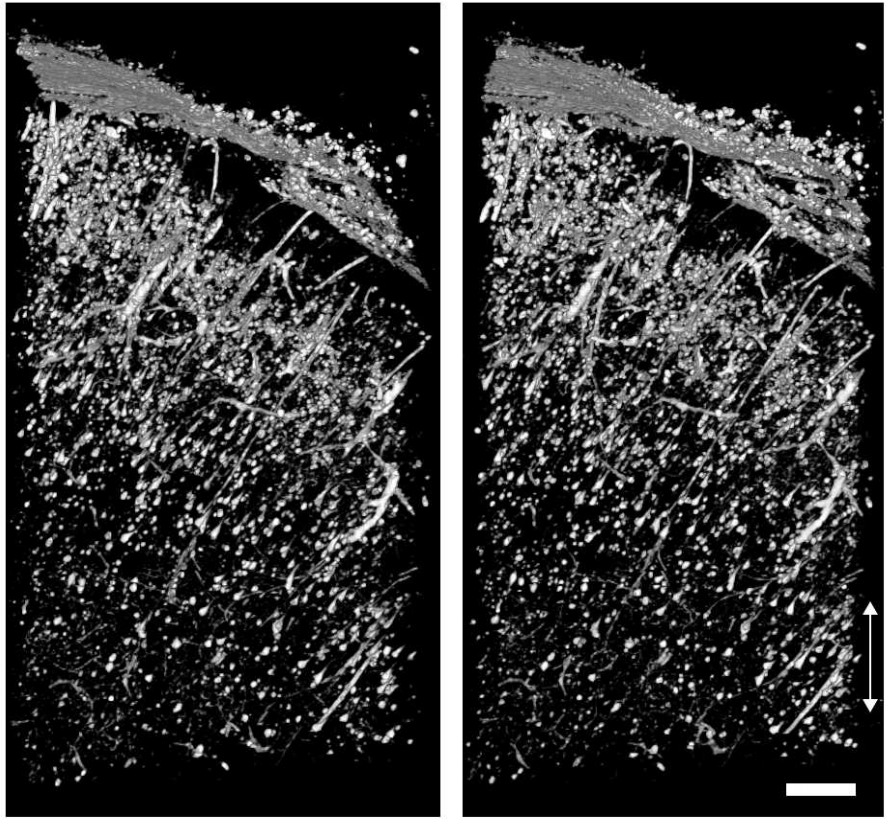

Fig. 2B (actual size)